\documentclass[preprint,prd,nofootinbib,preprintnumbers]{revtex4}
 \usepackage{graphicx,psfrag,epsfig,epsf,colordvi,slashed,amsmath,bm}

\newcommand{\lsim}{\mbox{\raisebox{-.6ex}{~$\stackrel{<}{\sim}$~}}}
\newcommand{\gsim}{\mbox{\raisebox{-.6ex}{~$\stackrel{>}{\sim}$~}}}

\begin{document}
\begin{flushright}
OSU-HEP-13-02\\
MAN/HEP/2012/017\\
UMD-PP-013-003 	
\end{flushright}
\title{Post--Sphaleron Baryogenesis and an Upper Limit on the \\[-0.07in]
Neutron--Antineutron Oscillation Time\\
\vspace{0.3cm}}

\author{{\bf K.S. Babu$^1$, P.S. Bhupal Dev$^2$, Elaine C.F.S. Fortes$^3$ and
R.N. Mohapatra$^4$}}
\affiliation{\vspace{2mm}
$^1$Department of Physics, Oklahoma State University,
Stillwater, OK 74078, USA\\
$^2$Consortium for Fundamental Physics, School of Physics and Astronomy,\\ University of Manchester, Manchester, M13 9PL,  United Kingdom\\
$^3$Instituto de F\'isica Te\'orica-Universidade Estadual Paulista, R. Dr. Bento Teobaldo Ferraz 271, S\~ao Paulo-SP, 01140-070, Brazil\\
$^4$Maryland Center for Fundamental Physics and
Department of Physics, University of Maryland, College Park, MD
20742, USA
}
\begin{abstract}

A recently proposed scenario for  baryogenesis, called post--sphaleron baryogenesis (PSB) is discussed
within a class of quark--lepton unified framework based on the gauge symmetry $SU(2)_L\times SU(2)_R\times SU(4)_c$ realized
in the multi--TeV scale.  The baryon asymmetry of the universe in this model is produced below the electroweak
phase transition temperature after the sphalerons have decoupled  from the Hubble expansion.
These models embed naturally the seesaw mechanism for neutrino masses, and predict color-sextet scalar particles
in the TeV range which may be accessible to the LHC experiments.  A necessary consequence of this scenario is the baryon number violating
$\Delta B=2$ process of neutron--antineutron ($n-\bar{n}$) oscillations. In this paper we show that the constraints of PSB, when combined
with the neutrino oscillation data and restrictions from flavor changing neutral currents mediated by the colored scalars
imply an upper limit on the $n-\bar{n}$ oscillation time of $5 \times 10^{10}$ {\it sec.} regardless of the quark--lepton unification scale. If
this scale is relatively low, in the $(200-250)$ TeV range, $\tau_{n-\bar{n}}$ is predicted to  be less than $10^{10}$ {\it sec.}, which is accessible to the next generation of proposed experiments.
\end{abstract}

\maketitle

\section{Introduction}

It is widely believed that understanding the origin of matter--antimatter asymmetry in the universe holds an important clue to  physics beyond the Standard Model (SM). A distinguishing signature of the nature of the new physics is the epoch at which baryogenesis occurs. In a series of recent papers~\cite{psb1a,psb1b,psb2} we have proposed and studied a new mechanism, termed post-sphaleron baryogenesis (PSB), where this dynamics occurs
at or below the TeV scale. This mechanism takes advantage of the baryon number violating decays of a new particle, either a scalar or a fermion,
which couples to the SM fermions through a higher-dimensional operator (with dimension $d\geq 9$).  If these decays go out of equilibrium
near the TeV scale, then the epoch of baryogenesis would be below the electroweak phase transition temperature, when the sphalerons have already decoupled due to the Hubble expansion of the universe.
The low baryogenesis scale arises if the process mediated by the higher-dimensional operator, $\cal O$, is in the observable range. This scenario is not only distinct from all other available baryogenesis mechanisms such as leptogenesis (see e.g., Ref.~\cite{Davidson:2008bu}) or electroweak baryogenesis (see e.g., Ref.~\cite{Morrissey:2012db}) but also involves TeV scale new particles accessible at the Large Hadron Collider (LHC) when an ultraviolet complete version of this theory is presented, and leads to interesting low energy phenomena accessible to non-accelerator searches as well.

A specific realization of the scenario proposed in Ref.~\cite{psb1a} is based on the gauge group $SU(2)_L\times SU(2)_R\times SU(4)_c$~\cite{ps} with a quark-lepton unified generalization~\cite{mm} of the seesaw mechanism~\cite{seesaw}  with TeV seesaw scale. The effective $d=9$ operator $\cal O$ in this model that couples to a TeV scale scalar field $S$ arises from the exchange of color-sextet fields.  These are part of the $SU(2)_R$ triplet Higgs field responsible for $B-L$ symmetry breaking and the seesaw mechanism.
In this model, the same operator $\cal O$ that leads to baryogenesis also leads to the baryon number violating process of neutron--antineutron  ($n-\bar{n}$) oscillation~\cite{nnb-review}. It is therefore natural to expect a connection between the amount of baryon asymmetry created
in the early universe and the strength of $n-\bar{n}$ oscillation amplitude. A realistic model of this type must reproduce the correct neutrino mass and mixing parameters, as measured by various neutrino oscillation experiments, and also satisfy the flavor changing neutral current (FCNC) constraints which arise in this case due to exchange of the color-sextet scalar fields. An investigation of these issues was initiated in Ref.~\cite{psb2}, where it was pointed out that if the color-sextet fields are in the TeV to sub-TeV range, consistence with FCNC constraints implies that neutrino masses must arise via a type--II seesaw mechanism and must exhibit an inverted mass hierarchy.  We presented a specific realization of this idea within a version~\cite{mm} of quark-lepton unified  $SU(2)_L\times SU(2)_R\times SU(4)_c$ model that embeds the type--II seesaw mechanism.  We also predicted the $n-\bar{n}$ oscillation to be sizable in this scenario if the model has to satisfy the constraints of generating adequate baryon asymmetry.
This model may also be testable via searches for the color-sextet scalar bosons at the LHC~\cite{color}.

We wish to point out that there have been other proposals for low-scale baryogenesis~\cite{hall,cline,others}. Our scenario differs from them not only in that we employ a model  that connects the new physics to neutrino masses but it also makes a specific testable prediction for a baryon number violating process of neutron-anti-neutron oscillation (as we show below) as well as new TeV scale particles at colliders. Furthermore, the mechanism for baryogenesis in our paper differs from those in Refs.~\cite{hall,cline,others} in two ways: (a) the operator responsible for baryogenesis in our case is different; (b) the one loop absorptive part that generates the primordial $C\!P$ asymmetry in our model involves flavor changing effects involving the $W$-exchange, whereas in the above papers it involves new fields beyond the SM.

While this paper is a follow-up to our earlier paper~\cite{psb2}, it presents several new results:
\begin{itemize}

\item   We present detailed constraints on the masses and couplings of the color-sextet scalar fields from various FCNC constraints. While Ref.~\cite{psb2} focussed on tree level constraints, here we include the one loop box diagram effects which provide stronger constraints on different flavor combinations of the sextet couplings.

\item We have found a one-loop $W$-exchange contribution to the $n-\bar{n}$ amplitude which gives an enhanced rate for $n-\bar{n}$ transition rate compared to 
the one given in Ref.~\cite{psb2}.

\item A striking new result of the present paper  is an absolute upper limit on the $n-\bar{n}$ oscillation time $\tau_{n-\bar{n}}$  of $5\times 10^{10}$ {\it sec.} irrespective of the $B-L$ breaking scale, which follows from the fact that we must generate enough baryon asymmetry via this mechanism. This oscillation time is within the accessible range for the next
generation of proposed searches for this process~\cite{proto}.

\end{itemize}

 The rest of this paper is organized as follows: In Section~\ref{review}, we review the basic features of our model. In Section~\ref{FCNC}, we summarize the FCNC constraints on the Yukawa couplings in our model; in Section~\ref{PSB}, we discuss various constraints that need to be satisfied in order to generate the observed baryon asymmetry using the PSB mechanism; and in Section~\ref{nnb}, we give the model predictions for $n-\bar{n}$ oscillation time and the resulting upper limit on it. Our conclusions are given in Section~\ref{con}.  In Appendix A, we present an explicit
 calculation of baryon asymmetry generated by using $B$--conserving vertices in a toy model.  This example shows the consistency of our baryon
 asymmetry generation mechanism using $W$ boson loops.

  \section{Review of the model}\label{review}

  We start by reviewing the basic features of our model~\cite{psb2}, based on
  the quark-lepton unified gauge group $SU(2)_L\times SU(2)_R\times SU(4)_c$  
  with SM fermions plus the right-handed neutrino belonging to $(2,1,4)\oplus (1,2,4) $ representations of the group in the well known left-right symmetric way~\cite{LRSM}. The Higgs sector of the model consists of  $(1,1,15)$, $(1,3,10)$, $(2,2,1)$ and $(2,2,15)$. The first stage of the symmetry breaking is implemented by a $(1,1,15)$ Higgs field which splits the $SU(4)_c$ scale $M_c$  from the remaining ones with $M_c\gsim 1400$ TeV \cite{wv} to satisfy the constraint from rare kaon decay: BR$(K_L^0\to \mu^\pm e^\mp)<4.7\times 10^{-12}$~\cite{Ambrose:1998us}. The surviving $SU(2)_L\times SU(2)_R\times U(1)_{B-L}\times SU(3)_c$ gauge symmetry is then broken in two stages down to the SM, i.e. by the Higgs field $(1,3,1)$ to the symmetry $SU(2)_L\times U(1)_{I_{3R}}\times U(1)_{B-L}$ which subsequently breaks down to the SM by the Higgs field $(1,3,\overline{10})$. The second stage is where the $B-L$ symmetry breaks down and the right-handed neutrinos acquire
mass by the usual seesaw mechanism~\cite{seesaw}. We denote this scale by $v_{BL}$, which is an essential parameter in
our discussion below.  It is also possible that the $(1,3,1)$ Higgs field is absent in the spectrum, in which case
  the  $SU(2)_L\times SU(2)_R\times U(1)_{B-L}$ gauge symmetry breaks directly down to the SM symmetry via the vacuum expectation
  value (vev) of the $(1,3,\overline{10})$ field. The SM Higgs field is a linear combination of the $(2,2,1)$ and $(2,2,15)$ Higgs fields.

  To discuss the mechanism for baryogenesis in the model, we first note that under $SU(2)_L \times U(1)_Y \times SU(3)_c$, the $(1,3,\overline{10})$ field, denoted by $\Delta$, decomposes as
 \begin{eqnarray}
	 \label{delta}
	 \Delta(1,3,\overline{10}) &=&  \Delta_{uu}(1, -\frac{8}{3}, 6^*)~\oplus~ \Delta_{u d}	(1,-\frac{2}{3}, 6^*)~\oplus~ \Delta_{d d}(1,+\frac{4}{3}, 6^*)
 ~\oplus~ \Delta_{u e}	(1,\frac{2}{3}, 3^*)\nonumber\\
&& ~\oplus~ \Delta_{u \nu}	(1,-\frac{4}{3}, 3^*)
 ~\oplus~ \Delta_{d e}	(1,\frac{8}{3}, 3^*)~\oplus~ \Delta_{d \nu}(1,\frac{2}{3}, 3^*)
 ~\oplus~ \Delta_{ee}	(1,{4}, 1)
 \nonumber \\
&& ~\oplus~ \Delta_{\nu e}	(1,{2}, 1) ~\oplus~ \Delta_{\nu \nu}	(1,0, 1)~.
 \end{eqnarray}
 The last field in the decomposition, $\Delta_{\nu \nu}(1,0, 1)$, is a neutral complex field whose real part acquires a vev $v_{BL}$ in the ground state  and can be written as
 $\Delta_{\nu \nu} =v_{BL}+\frac{1}{\sqrt{2}}(S+i\chi)$.  The field $\chi$ is absorbed by the $B-L$ gauge boson, while the
 real scalar $S$ remains as a physical Higgs particle.  It is the decay of this $S$ that will generate baryon asymmetry of the universe.
 The various color-sextet sub-multiplets of the field $\Delta (1,3,\overline{10})$ have couplings of the form
 \begin{eqnarray}
	 \label{lag}
{\cal  L}_I &	=	&\frac{f_{ij}}{2} \Delta_{dd} d_id_j + \frac{h_{ij}}{2} \Delta_{uu}u_iu_j + \frac{g_{ij}}{2\sqrt 2} \Delta_{ud}(u_id_j + u_jd_i) \nonumber\\
&&+\frac{\lambda}{2} \Delta_{\nu \nu}
\Delta_{dd}\Delta_{ud} \Delta_{ud}+{\lambda'} \Delta_{\nu \nu} \Delta_{uu}\Delta_{dd}\Delta_{dd} +{\rm h.c.}	
\label{Yuk}
 \end{eqnarray}
 Here the Yukawa couplings, as defined in Eq. (\ref{Yuk}), obey the boundary conditions $f_{ij} = h_{ij} = g_{ij}$ in the $SU(2)_L \times SU(2)_R \times  SU(4)_c$ symmetry limit.  All fermion fields here are right--handed, we have suppressed the chiral projection operators
 for simplicity.  There are analogous terms, dictated by left--right symmetry, where the left--handed fermion fields
 couple to the Higgs fields in the $(3,1,\overline{10})$ representation, with identical coupling strength as shown in Eq. (\ref{Yuk}). The last two terms in Eq. (\ref{Yuk}) are part of the Higgs potential, and are crucial for the generation
 of baryon asymmetry, with the boundary condition $\lambda' = \lambda$.  The color indices in these two terms are contracted by
 two $\epsilon_{ijk}$ factors.

Note that the $S$ field contained in $\Delta_{\nu \nu}$ is a real scalar field and therefore it can decay into both six quark and six anti-quark final states, thereby  violating baryon number by two units. The couplings of Eq. (\ref{Yuk}) allow for such baryon number violating decays of $S$.
If the right thermodynamic conditions are satisfied, it can generate baryon asymmetry in the presence of $C\!P$ violation. As shown in Ref.~\cite{psb2}, the CKM $C\!P$ violation is enough in this case although the presence of $C\!P$ violation in the $\Delta_{qq}$ couplings can help to enhance this.
The same interactions also generate a $d=9$ operator, once the vev of $S$ is inserted, that leads to neutron--antineutron oscillations.
In this paper, we argue that the right thermodynamic conditions are so restrictive that they imply $\tau_{n-\bar{n}}\leq 5\times 10^{10}$ {\it sec.} for arbitrary $v_{BL}$, and for low-scale $v_{BL}$ around 200 TeV, even more restrictive:  $\tau_{n-\bar{n}}\leq 10^{10}$ {\it sec.} which is accessible to the next generation $n-\bar{n}$ oscillation experiments~\cite{proto}. The significance of this result is that if in future experiments, the lower limit on $\tau_{n-\bar{n}}$ is found to exceed this limit, this model for PSB and neutrino masses will be ruled out.

\section{Restrictions of FCNC  on the model parameters} ~\label{FCNC}
It was noted in Ref. \cite{psb2} that tree-level exchange of color-sextet fields would result in new contributions
to  $\Delta F=2$ meson--antimeson
mixing, thereby yielding severe constraints on the masses and couplings of the color-sextet fields. Subsequently we have
realized that there are also important box diagrams which provide further constraints  coming both from $\Delta F=2$ meson--antimeson
mixing as well as flavor changing non-leptonic decays of $D$ and $B$ mesons.
In a forthcoming paper we shall present details of this analysis \cite{fortes}.  Here we summarize the main results,
which will be crucial in deriving the upper limit on $n-\bar{n}$ oscillation time within our model, consistent with the PSB mechanism.

\begin{figure}[htb]
\centering
\includegraphics[width=5cm]{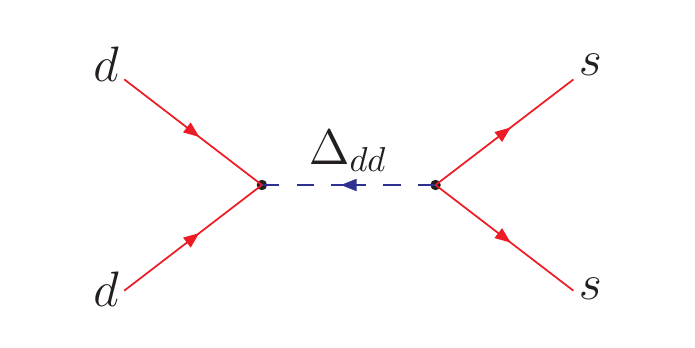}
\hspace{0.1cm}
\includegraphics[width=5cm]{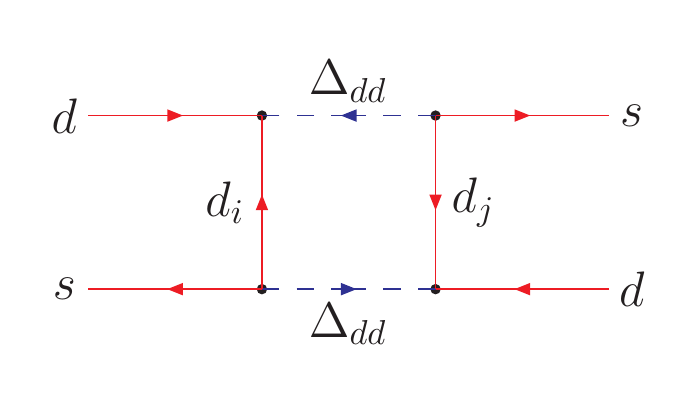}
\includegraphics[width=5cm]{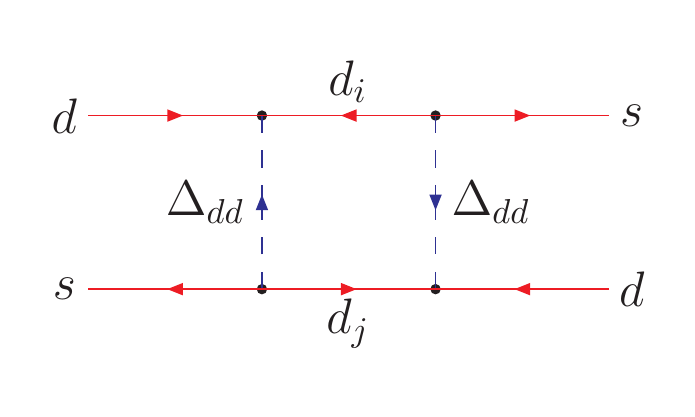}
\caption{Tree and box diagrams mediated by $\Delta_{dd}$ generating new contributions to
$K^0-\overline{K}^0$ mixing in the PSB model. Similar diagrams exist for $B^0-\overline{B}^0$ and $D^0-\overline{D}^0$ mixing, also involving the
exchange of $\Delta_{ud}$ and $\Delta_{uu}$ scalars.}
\label{fig-tree}
\end{figure}

Fig. \ref{fig-tree} illustrates new contributions to $K^0-\overline{K}^0$ mixing mediated by the $\Delta_{dd}$ color-sextet scalar field.
There are tree--level as well as box diagram contributions, which have different flavor structure.  Even if the tree--level diagram is
suppressed by choosing a specific flavor texture, the box diagram contributions can still provide strong constraints.  The effective $\Delta F = 2$ Hamiltonian
resulting from the $\Delta_{dd}$ exchange can be written as
\begin{eqnarray}\label{1d}
\mathcal{H}_{\Delta
F=2}&=&-\frac{1}{8}\frac{f_{i\ell}f_{kj}^{*}}{M_{\Delta_{dd}}^{2}}(\overline{d}_{kR}^{\alpha}\gamma_{\mu}d_{iR}^{\alpha})(\overline{d}_{j
R}^{\beta}\gamma^{\mu}d_{\ell R}^{\beta})
 +\frac{1}{256\pi^{2}}\frac{[(ff^{\dagger})_{ij}(ff^{\dagger})_{\ell k}+(ff^{\dagger})_{ik}(ff^{\dagger})_{\ell j}]}{M_{\Delta_{dd}}^{2}}\nonumber \\
 &&\times\left[(\overline{d}_{jR}^{\alpha}\gamma_{\mu}d_{iR}^{\alpha})(\overline{d}_{kR}^{\beta}\gamma^{\mu}d_{\ell R}^{\beta})+5(\overline{d}_{jR}^{\alpha}\gamma_{\mu}d_{iR}^{\beta})(\overline{d}_{kR}^{\beta}\gamma^{\mu}d_{\ell R}^{\alpha})\right]~.
 \end{eqnarray}
Here $i,j,k,\ell$ are flavor indices, while $\alpha, \beta$ are color indices.  The first term in Eq. (\ref{1d}) is from the tree-level
diagram, while the second term arises from the box diagram.  Setting  flavor indices $i=\ell=2$ and $j=k=1$ in Eq. (\ref{1d}) would
generate new contributions to $K^0-\overline{K}^0$ mixing.  There are analogous $\Delta F = 2$ FCNC contributions in the up--flavor
sector mediated by $\Delta_{uu}$ scalar for which the corresponding effective Hamiltonian can be obtained from Eq. (\ref{1d}) by replacing $d_i$
by $u_i$ and the coupling $f_{ij}$ by $h_{ij}$.  The constraint from $D^0- \overline{D}^0$ mixing will provide an important restriction
on the mass of $\Delta_{uu}$ in our analysis.

The effective $\Delta F = 2$ Hamiltonian resulting from the exchange of $\Delta_{ud}$ can be written as:
\begin{eqnarray}\label{1g}
    \mathcal{H}_{eff}&=&-\frac{1}{32}\frac{\widehat{g}_{ij}\widehat{g}_{kl}^{*}}{M_{\Delta_{ud}}^{2}}\left[(\overline{u}_{kR}^{\alpha}\gamma_{\mu}u_{iR}^{\alpha})(\overline{d}_{\ell R}^{\beta}\gamma^{\mu}d_{j R}^{\beta})+(\overline{u}_{kR}^{\alpha}\gamma_{\mu}d_{iR}^{\alpha})(\overline{d}_{\ell R}^{\beta}\gamma^{\mu}u_{j R}^{\beta})\right]\nonumber \\
    && +\frac{1}{256\pi^{2}} \frac{1}{64}\frac{1}{M_{\Delta_{ud}}^{2}}\left[(\widehat{g}\widehat{g}^{\dagger})_{ij}(\widehat{g}\widehat{g}^{\dagger})_{\ell k}+(\widehat{g}\widehat{g}^{\dagger})_{ik}(\widehat{g}\widehat{g}^{\dagger})_{\ell
    j}\right]\nonumber\\
   &&\times \left[(\overline{d}_{jR}^{\alpha}\gamma_{\mu}d_{iR}^{\alpha})(\overline{d}_{kR}^{\beta}\gamma^{\mu}d_{\ell R}^{\beta})+5(\overline{d}_{jR}^{\alpha}\gamma_{\mu}d_{iR}^{\beta})(\overline{d}_{kR}^{\beta}\gamma^{\mu}d_{\ell
   R}^{\alpha})\right]
  \end{eqnarray}
where we have defined $\widehat{g}_{ij}=(g_{ij}+g_{ji})/2$.

We apply standard methods to derive bounds on the couplings and masses of the color-sextet scalars from meson--antimeson mixing,
taking into account the renormalization of the effective four-fermion operator down to the meson mass scale, and using recent lattice
evaluation of the relevant matrix elements.  These constraints are listed in Table \ref{tab-mix}.
\begin{table}[h!]
\begin{tabular}{c|c|c} \hline\hline
Process & Diagram & Constraint on Couplings \\ \hline
 & Tree &  $|f_{22} f^*_{33} |\leq 7.04\times 10^{-4}\left(\frac{	M_{\Delta_{dd}}}{1~{\rm TeV}}\right)^2$\\
$\Delta m_{B_s}$ & Box & $\sum^3_{i=1}|f_{i3} f^*_{i2}|\leq 0.14 \left(\frac{M_{\Delta_{dd}}}{1~{\rm TeV}}\right)$\\
& Box & $\sum^3_{i=1}|\hat{g}_{i3} \hat{g}^*_{i2}|\leq 1.09 \left(\frac{M_{\Delta_{ud}}}{1~{\rm TeV}}\right)$ \\ \hline
& Tree & $ |f_{11} f^*_{33}|\leq 2.75\times 10^{-5}\left(\frac{	M_{\Delta_{dd}}}{1~{\rm TeV}}\right)^2$\\
$\Delta m_{B_d}$ & Box & $\sum^3_{i=1}|f_{i3} f^*_{i1}|\leq 0.03 \left(\frac{M_{\Delta_{dd}}}{1~{\rm TeV}}\right)$\\
& Box & $\sum^3_{i=1}|\hat{g}_{i3} \hat{g}^*_{i1}|\leq 0.21 \left(\frac{M_{\Delta_{ud}}}{1~{\rm TeV}}\right)$\\ \hline
& Tree & $ |f_{11} f^*_{22}|\leq 6.56\times 10^{-6}\left(\frac{M_{\Delta_{dd}}}{1~{\rm TeV}}\right)^2$\\
$\Delta m_{K}$ & Box & $\sum^3_{i=1}|f_{i2} f^*_{i1}|\leq 0.01 \left(\frac{M_{\Delta_{dd}}}{1~{\rm TeV}}\right)$\\
& Box & $\sum^3_{i=1}|\hat{g}_{i1} \hat{g}^*_{i2}|\leq 0.10 \left(\frac{M_{\Delta_{ud}}}{1~{\rm TeV}}\right)$\\ \hline
$\Delta m_{D}$ & Tree &  $|h_{11}h_{22}^*| \leq 3.72\times 10^{-6}\left(\frac{M_{\Delta_{uu}}}{1~{\rm TeV}}\right)^2$\\
& Box & $\sum^3_{i=1} |h_{i2}h_{i1}^*|\leq 0.01\left(\frac{M_{\Delta_{uu}}}{1~{\rm TeV}}\right)$ \\ \hline\hline
\end{tabular}
\caption{Constraints on the product of Yukawa couplings in the PSB model from $K^0-\overline{K}^0,~D^0-\overline{D}^0,~B^0_s-\overline{B}^0_s$ and $B^0_d-\overline{B}^0_d$ mixing.} \label{tab-mix}
\end{table}

\begin{figure}[htb]
\centering
\includegraphics[width=5cm]{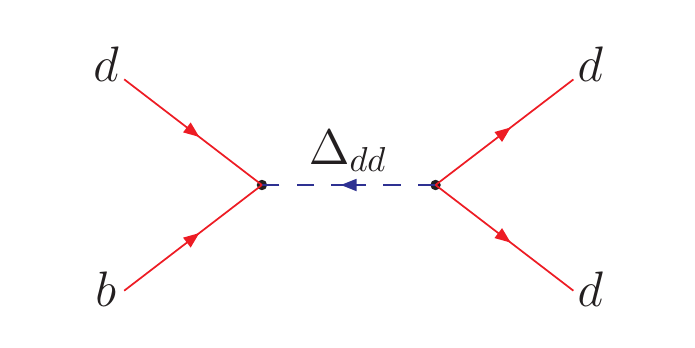}
\caption{Feynman diagram for $B$-decay mediated by the $\Delta_{dd}$-field in
the PSB model.}
\label{figB}
\end{figure}

The $\Delta F = 2$ effective Hamiltonian can also generate flavor changing non-leptonic decays of the type $B^- \rightarrow \phi \pi^-$
at the tree-level, mediated by $\Delta_{dd}$ scalar, via diagrams such as in Fig. \ref{figB}.
There are analogous diagrams mediated by $\Delta_{uu}$ and $\Delta_{ud}$ fields,
but we find that constraints from those diagrams are not so stringent, once $\Delta_{uu}$ field is assumed to be heavy,
as required by $D^0-\overline{D}^0$ mixing constraint.  In Table \ref{tab-bdecay} we present the various constraints arising from the $B$-meson decays.
These results are obtained by QCD factorization method \cite{fortes}. The numbers in the second column in Table~\ref{tab-bdecay} are to be multiplied by $(M_{\Delta_{dd}}/{\rm TeV})^2$.
\begin{table}[h!]
\begin{tabular}{c|c}\hline\hline
Decay  & constraints on couplings \\ \hline
$B^-\to \pi^0\pi^-$ & $|f_{13}f^*_{11}|\leq 0.73$  \\ \hline
$\overline{B}^0_d\to \phi\pi^0 $  &$|f_{23}f^*_{12}|\leq 0.05$ \\\hline
$B^-\to \phi\pi^-$ & $|f_{23}f^*_{12}|\leq 0.03$  \\ \hline
$\overline{B}^0_d\to \phi\overline{K}^0 $  &$|f_{23}f^*_{22}|\leq 0.33$ \\\hline
$B^-\to \phi K^-$ & $|f_{23}f^*_{22}|\leq 0.3$  \\ \hline
$\overline{B}^0_d\to \pi^0\pi^0 $  &$|f_{13}f^*_{11}|\leq 0.43$ \\\hline
$\overline{B}^0_d\to \overline{K}^0 K^0 $  &$|f_{23}f^*_{12}|\leq 0.26$ \\\hline
$\overline{B}^0_d\to K^0K^0 $  &$|f_{13}f^*_{22}|\leq 0.52$ \\\hline
$B^-\to K^0 K^-$ & $|f_{23}f^*_{12}|\leq 0.3$  \\ \hline
$B^-\to \overline{K}^0 K^-$ & $|f_{13}f^*_{22}|\leq 0.6$  \\ \hline
$\overline{B}^0_d\to \overline{K}^0\pi^0 $  &$|f_{13}f^*_{12}|\leq 0.31$ \\\hline
$B^-\to \pi^0K^-$ & $|f_{13}f^*_{12}|\leq 0.46$  \\ \hline
$B^-\to \pi^-\overline{K}^0$ & $|f_{13}f^*_{12}|\leq 1.26$  \\ \hline\hline
\end{tabular}\label{table1}
\caption{Constraints on the product of the $f$-couplings from non-leptonic rare $B$-meson decays.  These constraints are
obtained in the QCD factorization method. The numbers in the second column should be multiplied by a factor $(M_{\Delta_{dd}}/{\rm TeV})^2$.}
\label{tab-bdecay}
\end{table}

In addition to satisfying the FCNC constraints, the PSB
model should also explain consistently the observed neutrino mixing angles and
mass-squared differences (for a review, see e.g.,~\cite{PDG}).
The FCNC constraints listed in Tables~\ref{tab-mix} and \ref{tab-bdecay} fix
the form of the $f$-matrix in Eq.~(\ref{lag}) to  be~\cite{psb2}
\begin{eqnarray}
	\label{f}
f=\left(\begin{array}{ccc} 0 & 0.95 & 1 \\ 0.95 & 0 & 0.01\\
1 & 0.01 & 0. 06\end{array}\right) ~.
\end{eqnarray}
This is written in a basis where the down quark mass matrix is diagonal.
Since in this basis, we can take the neutrino mass matrix (in the type-II seesaw) to be proportional to the $f$ matrix, in the leading order prior to the contribution from charged leptons are included, the atmospheric mixing can be chosen near maximal but more importantly, the mass hierarchy is inverted~\cite{psb2}. Excellent fit to all neutrino oscillation data was obtained in Ref.~\cite{psb2} with this form of the mass matrix~\footnote{Note that the fit presented in Ref.~\cite{psb2} yielded a ``large" $\theta_{13}= 8^\circ$,  which is consistent with the recent measurements of this mixing angle at Daya Bay~\cite{daya} and RENO~\cite{reno} experiments.}.
We have not been able to find any way to get normal hierarchy for the neutrinos that is consistent with FCNC constraints of Table~\ref{tab-mix}. Note that the couplings $g$ and $h$ of $\Delta_{ud, uu}$ respectively are related to $f$ via quark mixing as
\begin{eqnarray}
g= U_{\rm CKM}f, ~~ h	=	U_{\rm CKM} fU^{\sf T}_{\rm CKM}, 
\end{eqnarray}
assuming that the right-handed mixing matrix is roughly similar to the left-handed CKM matrix (as is generally expected in left-right models), the constraints on $h$ and $g$ in Table~\ref{tab-mix} require us to take the following hierarchy among the $\Delta$ masses: $M_{\Delta_{ud}} \lsim M_{\Delta_{dd}} \ll M_{\Delta_{uu}}$, with $M_{\Delta_{ud}}\gsim 3$ TeV, $M_{\Delta_{dd}} \gsim 5 $ TeV and $M_{\Delta_{uu}} \gsim 200$ TeV as the lowest values. Of course one could argue that we could make the couplings smaller to allow for even lighter $\Delta$ masses. However, we will see in Section~\ref{PSB} that smaller couplings are disfavored by the
cosmological constraints required to generate the observed baryon asymmetry.

We also note that in our model there are new contributions to lepton flavor violating (LFV) processes e.g. $\mu\to 3e$ and $\mu\to e\gamma$ from the exchange of $\Delta_{ee}$ fields. Since the $\Delta_{ee}$ fields in our model are assumed to be very heavy with mass of order of 100 TeV, the LFV constraints are easily satisfied. 

\section{Constraints of post-sphaleron baryogenesis}~\label{PSB}
 An important point to note is that if the diquarks $\Delta_{qq}$ have masses in the TeV range as discussed above, they will lead to a large rate for the baryon violating processes. As a result, the associated baryon violating processes e.g. $NN\to \pi$'s, $n-\bar{n}$ oscillation etc will remain in equilibrium till near the TeV scale and erase any pre-existing matter-antimatter asymmetry in the universe. So in this model, one must necessarily have a new mechanism for generating baryon excess below the electroweak phase transition temperature. Here we focus on the post-sphaleron baryogenesis~\cite{psb1a}, which is connected in our model to two popular ideas, i.e., seesaw for neutrino masses~\cite{seesaw}
and unification of quarks with leptons~\cite{ps}.

\begin{figure}[h!]
	\centering
	\includegraphics[width=7cm]{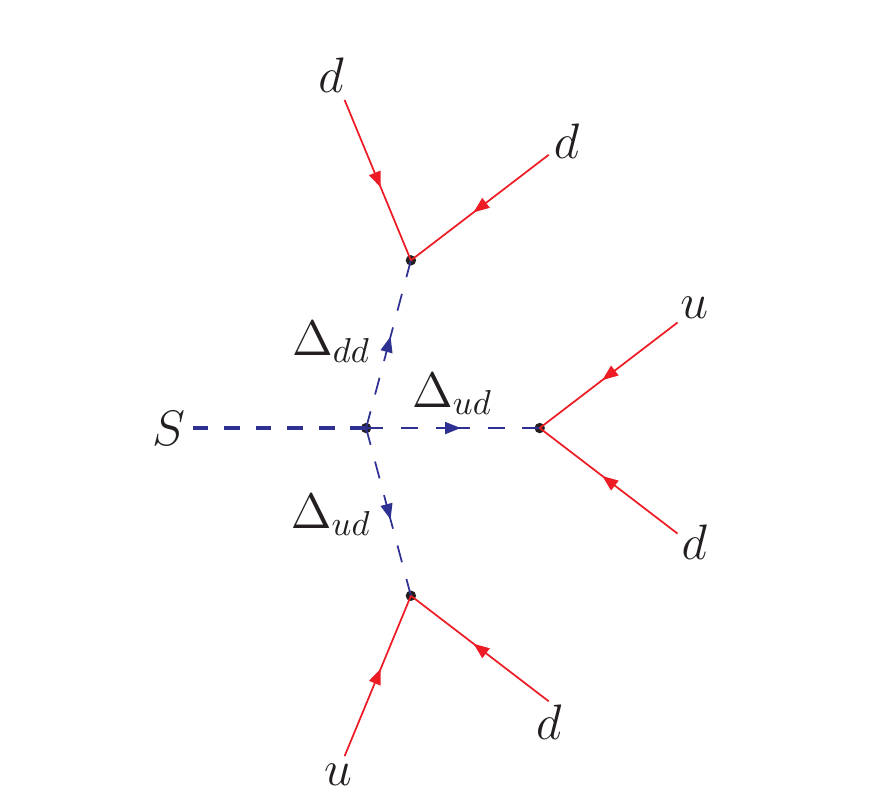}
	\caption{Tree-level diagram contributing to the decay $S\to 6q$ in the PSB model. A similar diagram for $S\to 6\bar{q}$ (which is possible since $S$ is a real scalar field) can be obtained by reversing the arrows of the quark fields. }
\label{fig6q}
\end{figure}

For any baryogenesis mechanism to be successful, all the three Sakharov's conditions~\cite{sakha} must be satisfied, and it turns out that in our case, due to the structure of the theory, some extra conditions outlined below must also be satisfied
by the model parameters. To understand the cosmological constraints, let us first outline the baryogenesis scenario: We assume that the $S$ field is the lightest member of the $(1,3,\overline{10})$ multiplet, i.e. it is lighter than the $\Delta_{qq}$ fields (so that it cannot have baryon-number conserving decays involving an on-shell $\Delta_{qq}$). It will go out of equilibrium and then decay after the electroweak phase transition. In this decay, it will produce six quarks and six anti-quarks (as shown in Figure~\ref{fig6q}) asymmetrically thereby creating the baryon excess. In our scenario, at some epoch when the universe is at a temperature $T \leq M_{\Delta_{ud, dd}}$ and $T \geq M_S$, the $S$-particle decay rate drops as a high power of $T^{13}$ and will go out of equilibrium. Then $S$-particles will simply ``drift'' along till $T\sim M_S$. At this epoch, its decay rate does not go down with temperature but remains frozen at  its value as if the $S$-particle were at rest. However, since the expansion rate of the universe is going down as $T^2$, at some temperature $T_d$, $H(T_d)\sim \Gamma_S$ and the $S$-particle will start decaying. In the post-sphaleron baryogenesis scenario, we must have $T_d \leq 100$ GeV so that the electroweak sphalerons have gone out of thermal
equilibrium (hence the name ``Post-sphaleron").  $T_d > 200$ MeV (the QCD phase transition temperature) must also be met, otherwise the success of nucleosynthesis will be
spoiled.

Let us now write down the constraints derived from this PSB mechanism on our
model:

\noindent{ \bf \underline {Condition I}:}~~
The decays of the $S$ field to quarks and anti-quarks are mediated by the exchange of
virtual $\Delta_{qq}$ fields.
The first condition to be satisfied for baryogenesis is that the $S\to 6 q$ decay rate must be smaller than the Hubble rate at some temperature near the electroweak phase transition epoch, i.e. $\Gamma_{S\to 6q} \leq H(T_{\rm ew})$. The $S$ fields then should drift around  till $T \leq T_{\rm ew}$ (which we will take for simplicity to be 100 GeV) and then they will decay; but we require them to decay before the QCD phase transition epoch which occurs around 200 MeV. If we denote this decay temperature as $T_d$, then the condition for PSB is $100$~GeV  $\geq T_d \geq 200$ MeV. To get $T_d$, we equate the decay rate $\Gamma_{S\to 6 q}$ to the Hubble rate $H(T_d)\simeq 1.66 g^{1/2}_*\frac{T^2_d}{M_{\rm Pl}}$, where $g_*$ is the number of relativistic degrees of freedom at $T_d$ and $M_{\rm Pl}=1.2\times 10^{19}$ GeV is the Planck mass. Using the Lagrangian of Eq. (\ref{lag}) and the mass hierarchy $M_{ud,dd}\ll M_{uu}$, we can estimate the dominant contribution to the
six-quark decay. This needs a careful counting of the final states, which we have carried out below.

We can write down the decay width as a product of the amplitude times the phase space factor for a six quark final state, and is given by
\begin{eqnarray}
\Gamma_S\equiv \Gamma(S\to 6 q)+\Gamma(S\to 6\bar{q}) =\frac{P}{\pi^9\cdot 2^{25}\cdot45}\frac{12}{4} |\lambda|^2 {\rm Tr}(f^\dagger f)[{\rm Tr}(\hat{g}^\dagger\hat{g})]^2 \left(\frac{M^{13}_S}{M^8_{\Delta_{ud}}M^4_{\Delta_{dd}}}\right)
\label{decay}
\end{eqnarray}
where the first term on the RHS is the 6-body phase space factor (for a constant matrix element) \cite{delborgo}, the factor 12 comes from counting the number of final states with different
$SU(3)_c$ color combinations, the factor 1/4 is due to the normalization of the coupling $g$ in terms of $\hat{g}$, and $P$ is a phase space integral done numerically. There is a $1/\sqrt{2}$ coming
from the S quartic vertex in the amplitude, and so there is a factor
1/2 in the rate.  This is compensated by the factor 2 obtained by
adding the two conjugate decay modes.The value of $P$ does not change much as a function of the mass ratios, for eg., we get the following two typical values:
\begin{eqnarray}
P = \left\{\begin{array}{cc}
1.13\times 10^{-4} & (M_{\Delta_{ud}}/M_S, M_{\Delta_{dd}}/M_S\gg1)\\
1.29\times 10^{-4}  &(M_{\Delta_{ud}}/M_S= M_{\Delta_{dd}}/M_S=2)~.
\end{array}\right.
\end{eqnarray}
We use the expression in Eq. (\ref{decay}) for $\Gamma_S$ and equate it to the Hubble rate $H(T_d)$ to evaluate $T_d$  which must be between $0.2 - 100$ GeV for successful PSB. Also as we will see below,  given a value of $T_d$, the dilution factor will constrain the value of $M_S$ which goes into the evaluation of the amount of baryon asymmetry as well as the the value of $T_d$ from decay width of $S$.

\noindent{\bf \underline{Condition II}:}
The second condition is that at the epoch of decay, the rate to six quarks must exceed other possible decay modes of $S$ such as $Zf\bar{f}, e\tau$ etc. This issue was analyzed in great detail in~\cite{psb2} and it was pointed out that for  $v_{BL}\lsim 100$ TeV, it implies an upper limit on $M_S \lsim 1$ TeV.
The condition I then implies that the masses of the color-sextet $\Delta$ fields should not be more than $5 - 10$ TeV, otherwise $T_d$ quickly falls below the lower bound of 0.2 GeV due to the high inverse power dependence on $M_{\Delta}$.
Note however that for larger $v_{BL}$, this condition is easily satisfied since the $S\to 6q$ decay rate which is independent of $v_{BL}$ dominates over the
other decays of $S$ which usually have a $1/v_{BL}^2$ dependence~\cite{psb2}.

\begin{figure}[h!]
	\centering
	\includegraphics[width=5cm]{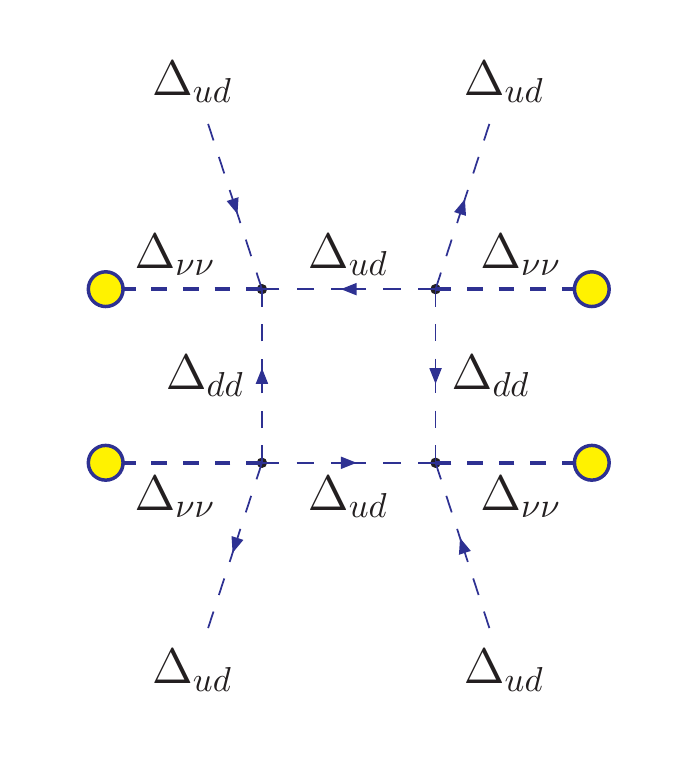}
	\caption{The box-diagram giving rise to the effective scalar quartic interaction terms.}
\label{fig4}
\end{figure}

\noindent{\bf \underline{Condition III}:}
A third condition arises from a field theoretic requirement of vacuum preserving color. The point is that the cubic term in the $\Delta$ fields in Eq.~(\ref{lag}), induced after the $\Delta_{\nu\nu}$ field acquires a vev, leads to effective potential terms of the form $-\frac{1}{16\pi^2}\left(\frac{\lambda v_{BL}}{M_\Delta}\right)^4(\Delta^\dagger\Delta)^2$~\cite{babu} via one-loop box graphs 
of the kind shown in Fig.~\ref{fig4}. To give the form of the effective potential, let us first write down the form of the potential $V_{BL}$ that leads to $B$-violation after $B-L$ symmetry breaking:
\begin{eqnarray}
V_{BL}  = \lambda \Delta_{\nu\nu}\left[\frac{1}{2}	\Delta^{i\alpha}_{ud} \Delta^{j\beta}_{ud}  \Delta^{k\gamma}_{dd} \epsilon_{ijk}\epsilon_{\alpha\beta\gamma}	+ 2\cdot\frac{1}{2} \Delta^{i\alpha}_{dd} \Delta^{j\beta}_{dd}  \Delta^{k\gamma}_{uu} \epsilon_{ijk}\epsilon_{\alpha\beta\gamma}+...\right]
\end{eqnarray}
where $i,j,k,\alpha,\beta,\gamma$ are all color indices. The $\Delta$ fields are the same as those in Eq.~(\ref{lag}) with color indices explicitly shown. This, after symmetry breaking, will generate via scalar box diagrams quartic terms for the $\Delta_{ud}$ field. The box diagram contributions to the effective potential as shown in Fig.~\ref{fig4}  can be written down as
\begin{eqnarray}
 V^{\rm 1-loop}_{\rm eff} = \frac{\alpha_1}{2}[{\rm Tr}(\Delta^\dagger_{ud}\Delta_{ud})]^2 + \frac{\alpha_2}{2}[{\rm Tr} (\Delta^\dagger_{ud}\Delta_{ud})^2]
 \end{eqnarray}
where
$$\alpha_1= -\frac{1}{8\pi^2}\frac{(\lambda v_{BL})^4}{(M^2_{\Delta_{ud}}-M^2_{\Delta_{dd}})^2}\left[\left(\frac{M^2_{\Delta_{ud}}+M^2_{\Delta_{dd}}}{M^2_{\Delta_{ud}}-M^2_{\Delta_{dd}}}\right)\ln\left(\frac{M^2_{\Delta_{ud}}}{M^2_{\Delta_{dd}}}\right) -2\right]$$
and $\alpha_2=-\frac{\alpha_1}{4}$.
Note that roughly for $v_{BL} \geq \frac{2\sqrt{\pi}M_\Delta}{\lambda}$, these effective terms will lead to vacuum instability along the $\Delta$ field direction, and therefore viewed naively, will be unacceptable. This would imply that the value of the $v_{BL}$ cannot be arbitrarily large for given masses of the $\Delta$ fields which are also constrained by the $T_d$ condition above given the mass of the $S$
field. We find that $\lambda v_{BL}$ cannot exceed the masses
of $\Delta_{ud}, \Delta_{dd}$ by more than a factor of $2-3$.

\noindent{\bf \underline{Condition IV}:} The final question one may ask is: could one allow very large values for $M_S$ so that proportionately larger
$M_\Delta$ values will lead to $T_d$ still being in the desirable range?
There is however one problem with this possibility, i.e. for large $M_S$,
the condition that the $S$ particle starts to decay below 100 GeV implies a
dilution factor that makes the net surviving baryon asymmetry too small. To see this, note that the dilution factor $d$ is given by the ratio of the entropy before and after decay~\cite{st}:
\begin{eqnarray}
d\equiv \frac{s_{\rm before}}{s_{\rm after}}\simeq \frac{g^{-1/4}_*0.6(\Gamma_S M_{\rm Pl})^{1/2}}{rM_S}
\end{eqnarray}
where $r=\frac{n_S}{s}$ at the epoch of decay. This dilution factor is roughly
estimated to be $\sim \frac{T_d}{M_S}$.

\begin{figure}[h!]
	\centering
	\includegraphics[width=5cm]{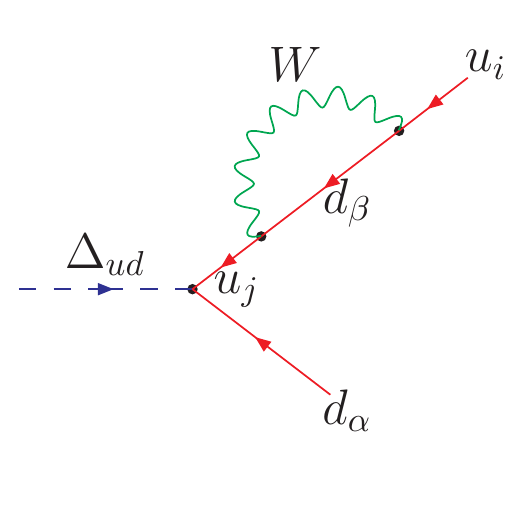}
\hspace{1cm}
	\includegraphics[width=5cm]{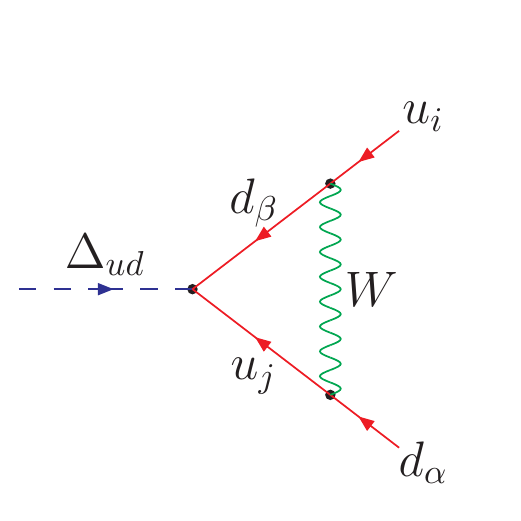}
	\caption{The wave function and vertex correction contribution to the $C\!P$ asymmetry in our PSB model.}
\label{fig:asy}
\end{figure}

On the other hand, a calculation of the primordial $C\!P$ asymmetry gives
\begin{eqnarray}
	\epsilon_{\rm wave} &\simeq& \frac{g^2}{64\pi {\rm  Tr}(f^\dagger f)}f_{j\alpha}V_{j\beta}
	V^*_{i\beta}f_{i\alpha}\delta_{i3}\frac{m_tm_j}{m_t^2-m_j^2}
\sqrt{\left(1-\frac{m_W^2}{m_t^2}+\frac{m_\beta^2}
	{m_t^2}\right)^2-4\frac{m_\beta^2}{m_t^2}}\nonumber\\
	&&
	\left[2\left(1-\frac{m_W^2}{m_t^2}+\frac{m_\beta^2}{m_t^2}\right)
	+\left(1+\frac{m_\beta^2}{m_t^2}\right)
	\left(\frac{m_t^2}{m_W^2}+\frac{m_\beta^2}{m_W^2}-1\right)
	-4\frac{m_\beta^2}{m_W^2}\right], \label{asy1}\\
\epsilon_{\rm vertex} &\simeq & \frac{g^2}{32\pi {\rm Tr}(f^\dagger f)}f_{j\beta}V^*_{i\beta}
	V_{j\alpha}f_{i\alpha}\delta_{i3}\frac{m_jm_\beta}{m_W^2}\left[1+\frac{3m_W^2}{2\langle p_1\cdot p_2\rangle}
	\ln\left(1+\frac{2\langle p_1\cdot p_2\rangle}{m_W^2}\right)\right] \label{asy2}
\end{eqnarray}
for the wave function and vertex correction diagrams respectively, as shown in Fig.~\ref{fig:asy}. Here $\langle p_1\cdot p_2 \rangle$ denotes the thermal average over the scalar product of the external momenta of the two quarks, which is of order $M_S^2/6$.  Our calculation is done in the unitary gauge and there are no other contributions to the primordial $C\!P$ asymmetry, that can cancel this contribution.
Note that we require one of the external legs to be the top-quark in order to get a non-zero absorptive part. Numerically, the vertex term turns out to be the dominant one with $\epsilon \sim 10^{-8}$ or so in this particular realization of PSB. This means that the dilution factor must not be less than about 1\%, or in other words, $M_S$ must be smaller than 10 TeV (since $T_d\leq 100$ GeV), in order to explain the observed baryon asymmetry, $\eta_B\equiv(n_b-n_{\bar b})/n_\gamma=(6.04\pm 0.08)\times 10^{-10}$~\cite{Ade:2013lta}.

It is important to note here that the loop diagrams in Fig.~\ref{fig:asy} giving rise to a non-zero $C\!P$-asymmetry do not involve baryon-number violating interactions. This point is further clarified in Appendix~\ref{appx}. 

\section{Prediction for $\tau_{n-\bar{n}}$}~\label{nnb}
\begin{figure}[htb]
	\centering
	\includegraphics[width=7cm]{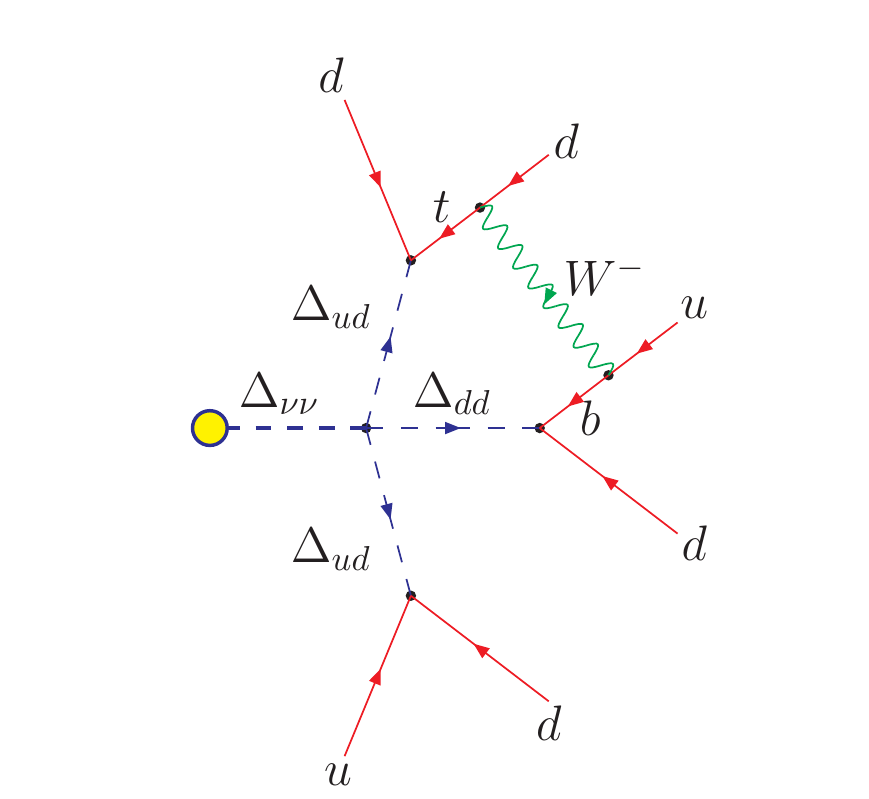}
	\caption{One-loop contribution to the $n-\bar{n}$ amplitude in the PSB model.}
\label{fig:loop}
\end{figure}

We now present the model predictions for the $n-\bar{n}$ oscillation time. We will show that under the constraints of PSB on the model parameters as discussed above, there is an absolute upper bound on the $\tau_{n-\bar{n}}$. To understand this, we first note that the $n-\bar{n}$ oscillation (or  the $\Delta B=2$ amplitude) in our model arises from the exchange of three color-sextet $\Delta$ fields. There are two generic contributions which have the form:
\begin{eqnarray}
A^{\rm tree}_{n-\bar{n}}\simeq \frac{f_{11}g^2_{11}\lambda v_{BL}}{M^2_{\Delta_{dd}}M^4_{\Delta_{ud}}}+\frac{f^2_{11}h_{11}\lambda' v_{BL}}{M^4_{\Delta_{dd}}M^2_{\Delta_{uu}}}~.
\end{eqnarray}
Note that both terms involve the coupling $f_{11}$. But a look at Eq.~(\ref{f}) tells us that at the tree-level, this coupling has to be vanishingly small to satisfy the FCNC constraints. However, the choice of $f$ matrix in Eq.~(\ref{f}) is not unique and we could as well choose a very small value for $f_{11}$ (e.g. $\lsim 10^{-6}$) without affecting the FCNC constraints. One would then think that the $n-\bar{n}$ amplitude could be as small as one wants. However,  there is
an one-loop diagram as shown in Figure~\ref{fig:loop} that sets a lower bound on the value of $f_{11}$.
The contribution of the one-loop diagram to $n-\bar{n}$ amplitude is given by
\begin{eqnarray}
	A^{\rm 1-loop}_{n-\bar{n}} &\simeq& \frac{g^2g_{11}g_{13}f_{13}V^*_{ub}V_{td}\lambda v_{BL}}
	{128 \pi^2M^2_{\Delta_{ud}}}\left(\frac{m_t m_b}{m^2_W}\right)F \langle \bar{n}|{\cal O}^2_{RLR}|n\rangle
\label{1loop}
\end{eqnarray}
where the one-loop function is given by
\begin{eqnarray}
F&=&\frac{1}{M^2_{\Delta_{ud}}-M^2_{\Delta_{dd}}}\left[\frac{1}{M^2_{\Delta_{ud}}}\ln\left(\frac{M^2_{\Delta_{ud}}}{m^2_W}\right)-\frac{1}{M^2_{\Delta_{dd}}}\ln \left(\frac{M^2_{\Delta_{dd}}}{m^2_W}\right)\right]\nonumber\\
&& +\frac{1}{M^2_{\Delta_{ud}}M^2_{\Delta_{dd}}}\frac{1-(m^2_t/4m^2_W)}{1-(m^2_t/m^2_W)}\ln\left(\frac{m^2_t}{m^2_W}\right),
\label{loop-func}
\end{eqnarray}
and the operator ${\cal O}^2_{RLR}$ is given by
\begin{eqnarray}
{\cal O}^2_{RLR} =  (u^{\sf T}_{iR}Cd_{jR})(u^{\sf T}_{kL}Cd_{l L})(d^{\sf T}_{m R}Cd_{nR}) \Gamma^s_{ijklmn},
\label{ORLR}
\end{eqnarray}
with $\Gamma^s_{ijklmn}=\epsilon_{mik}\epsilon_{njl}+\epsilon_{nik}\epsilon_{mjl}+\epsilon_{mjk}\epsilon_{nil}+\epsilon_{njk}\epsilon_{mil}$,
 where we have used the notation in Ref. \cite{bag}. The matrix element of this operator between the $n$ and $\bar{n}$ states has been evaluated in the MIT bag model in Ref.~\cite{bag}, and we take their fit A value:
\begin{eqnarray}
\langle \bar{n}|{\cal O}^2_{RLR}|n\rangle =-0.314\times 10^{-5}~{\rm GeV}^6
\end{eqnarray}
to predict the upper bound on $\tau_{n-\bar{n}}$ in our model. Note that in the last term of Eq.~(\ref{loop-func}), the factor $(1 -m^2_t /4m^2_W )$ is nearly zero since $m_t \simeq 2m_W$ . This factor arises from including the longitudinal components of W boson in the evaluation of the diagram.
Here the approximation $M^2_{\Delta_{ud,dd}}\gg m^2_t , m^2_W$	has been made. Also, a Fierz transformation has been made to obtain the operator in Eq.~(\ref{ORLR}) in the scalar form shown here.

The $n-\bar{n}$ amplitude in Eq.~(\ref{1loop}) can be translated into the $n-\bar{n}$ oscillation time as follows:
\begin{eqnarray}
\tau_{n-\bar{n}}^{-1} \equiv \delta m = c_{\rm QCD}(\mu_\Delta, 1~{\rm GeV})\left|A^{\rm 1-loop}_{n-\bar{n}}\right|,
\end{eqnarray}
where $c_{QCD}$ is the RG running factor in bringing down the amplitude~(\ref{1loop}) originally evaluated at the $\Delta$-scale to the neutron scale~\cite{Winslow}:
\begin{eqnarray}
c_{\rm QCD}(\mu_\Delta,1{\rm GeV}) & = &
\left[\frac{\alpha_s(\mu_\Delta^2)}{\alpha_s(m_t^2)}\right]^{8/7}
\left[\frac{\alpha_s(m_t^2)}{\alpha_s(m_b^2)}\right]^{24/23}
\left[\frac{\alpha_s(m_b^2)}{\alpha_s(m_c^2)}\right]^{24/25}
\left[\frac{\alpha_s(m_c^2)}{\alpha_s(1~{\rm GeV}^2)}\right]^{8/9}.
\label{cqcd}
\end{eqnarray}
Here we have assumed $\mu_\Delta$ to be the geometric mean of $M_{\Delta_{ud}}$ and $M_{\Delta_{dd}}$, and have used $\mu_\Delta\sim {\cal O}({\rm TeV})$
to obtain $c_{\rm QCD}\simeq 0.18$.

\begin{figure}[tb]
\includegraphics[width=7cm]{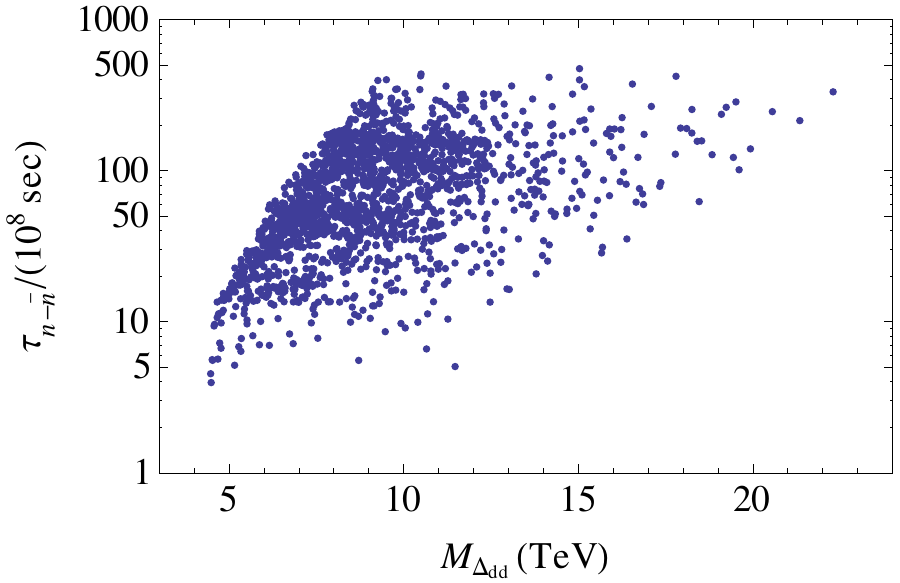}
\hspace{1cm}
\includegraphics[width=7cm]{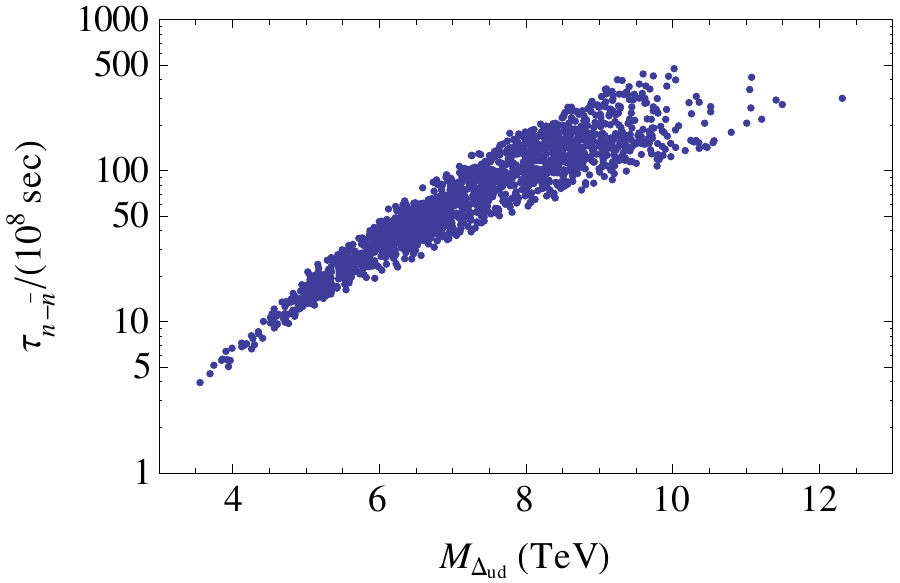}
\caption{Scatter plots for $\tau_{n-\bar{n}}$ as a function of the $\Delta$
masses $M_{\Delta_{ud}}, M_{\Delta_{dd}}$.}
\label{fig:scat1}
\end{figure}
\begin{figure}[tb]
\includegraphics[width=7cm]{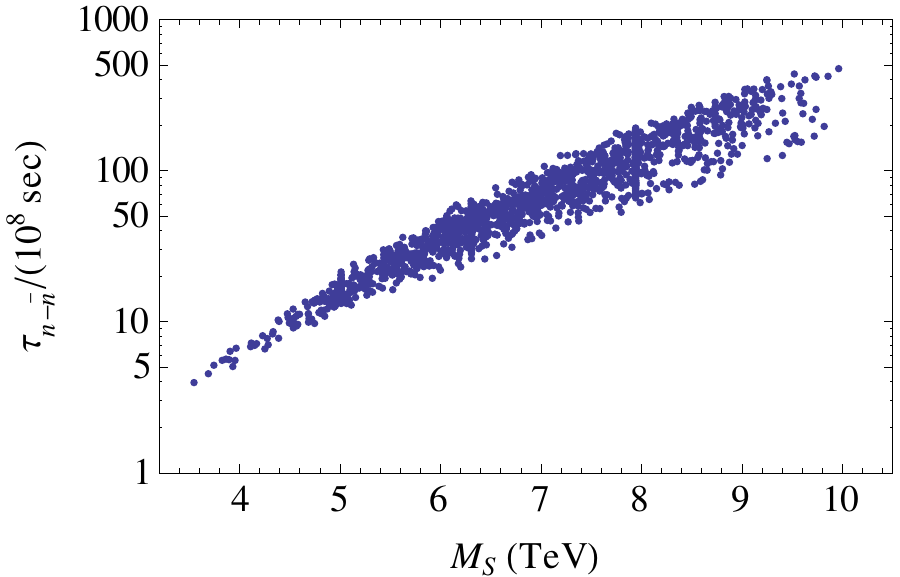}
\hspace{1cm}
\includegraphics[width=7cm]{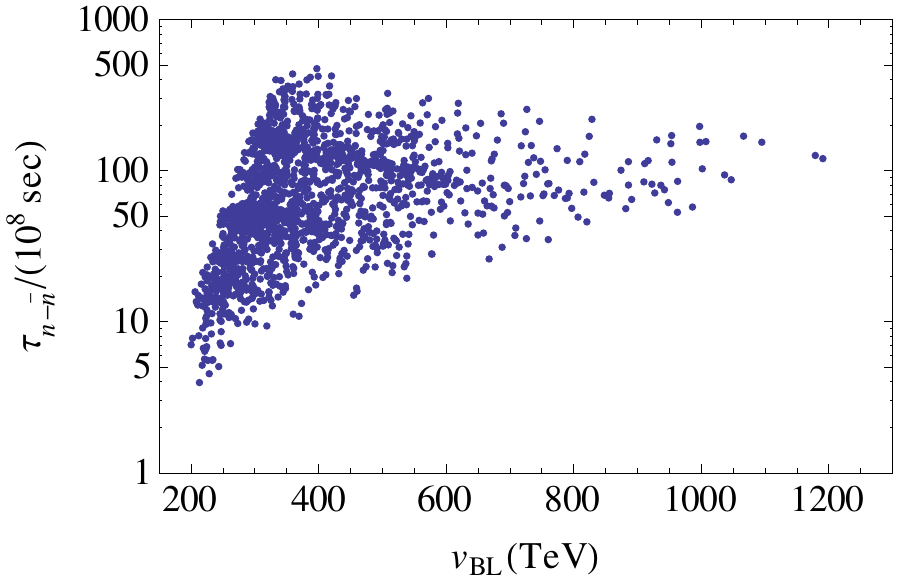}
\caption{Scatter plots for $\tau_{n-\bar{n}}$ as a function of the real scalar
mass $M_S$ and the $B-L$ breaking scale $v_{B-L}$.}
\label{fig:scat2}
\end{figure}
Using all the PSB constraints described in the previous section, we vary all
the model parameters in the allowed range. In particular, we perform a numerical scan (with logarithmic scale) over the mass parameter $M_S$ between 100 GeV and 10 TeV, the $B-L$ breaking scale $v_{BL}$ from 10 TeV upwards, and the masses $M_{\Delta_{ud,dd}}$ between $M_S$ and $v_{BL}$. We also vary the coupling $\lambda$ (the allowed values were found to be between $0.01-1$) as well as the overall scale in the $f$-matrix given by Eq.~(\ref{f}) (its allowed values were
between $0.5-1.6$).

\begin{figure}[tb]
\centering
\includegraphics[width=7cm]{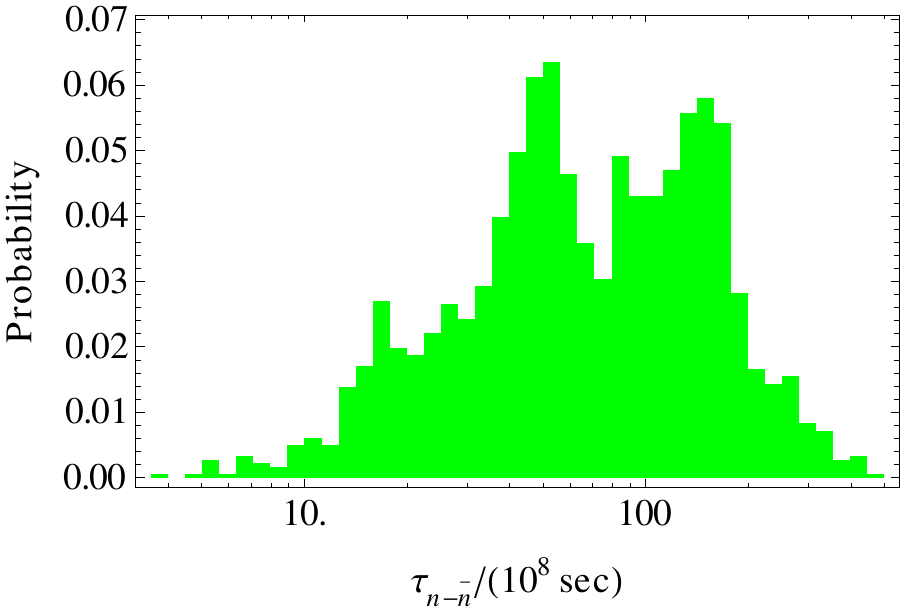}
\caption{The likelihood probability for a particular value of  $\tau_{n-\bar{n}}$ as given by the model parameters.}
\label{fig:scat3}
\end{figure}

We obtain an absolute upper limit on the
oscillation time of $\tau_{n-\bar{n}}\leq 4.7\times 10^{10}$ {\it sec.}.
This is demonstrated in Figures~\ref{fig:scat1} and \ref{fig:scat2} for the most relevant model parameters, namely $v_{BL},M_{\Delta}$ and $M_S$. A probability distribution of the predictions for $\tau_{n-\bar{n}}$ is shown in Figure~\ref{fig:scat3}. Note that the current experimental lower limit is $\tau_{n-\bar{n}}^{\rm expt}\geq 3.5\times 10^8$ {\it sec.}~\cite{kamio}. We further note that our predicted upper limit on $\tau_{n-\bar{n}}$ gets even stronger for low $B-L$ scale, e.g., for $v_{BL}$ around 200 TeV, $\tau_{n-\bar{n}}\lsim 10^{10}$ {\it sec.}, which is within reach of the proposed $n-\bar{n}$ oscillation experiments~\cite{proto}. Note that for $v_{BL}\lsim 200$ TeV, there are no allowed points in our model since the $S\to 6q$ decay rate no longer remains the dominant decay mode while satisfying all the other constraints discussed in the previous two sections.

\section{Conclusion}~\label{con}
We have presented the predictions for neutron-anti-neutron oscillation in a
new low-scale baryogenesis scenario, namely the post-sphaleron baryogenesis. We find that the requirements of successful baryogenesis, together with the flavor changing neutral current constraints, restrict the model parameter space significantly to give an absolute upper limit on $\tau_{n-\bar{n}}\leq 5\times 10^{10}$ {\it sec.}, which is independent of the $B-L$ breaking scale. For a low $B-L$ scale around 200 TeV, the upper limit is even stronger: $\tau_{n-\bar{n}}\leq 10^{10}$ {\it sec.}, a value in the range accessible to the future round of $n-\bar{n}$  searches. Interestingly, this model also allows a realistic neutrino masses and mixing observed although it is consistent only with inverted mass hierarchy pattern. Thus evidence for normal mass hierarchy will rule out this scenario. We hope this result will strengthen the theoretical and experimental motivations for dedicated searches for neutron-anti-neutron oscillation searches in near future.
\begin{acknowledgments}
We thank  Geoff Greene, Yuri Kamyshkov, Chris Quigg, Mike Snow and Albert Young for discussions and encouragement. We also thank Alexander Khanov and Xiao-Gang He for useful comments. The work of KSB is supported in part by the US Department of Energy Grant No. DE-FG02-04ER41306, PSBD is supported by the Lancaster-Manchester-Sheffield Consortium for Fundamental Physics under STFC grant ST/J000418/1, ECFSF is supported by FAPESP under contract No. 2011/21945-8, and RNM is supported
by National Science Foundation grant No. PHY-0968854. 
\end{acknowledgments}


\appendix
\section{Baryon Asymmetry Calculation in a Toy Model}\label{appx}
In this Appendix, we discuss whether a theorem discussed by Nanopoulos and Weinberg \cite{nanop} (NW) regarding the nature of one loop contribution that can lead to nonzero baryon asymmetry $\epsilon_B$, applies to our model. According to this theorem, if $\alpha_B$ is the strength of the baryon number violating coupling, non-zero baryon asymmetry can arise from one loop contributions that involve only the  $B$-violating interactions, i.e., $\epsilon_B\propto \alpha^3_B$, with two powers of $\alpha_B$ coming from tree amplitudes and one power from the loop contribution. On the other hand, an explicit calculation in our model shows that indeed a non-zero $\epsilon_B$ can arise in order $\alpha^2_B$, as shown in Eqs.~(\ref{asy1}) and (\ref{asy2}). We pointed out in Ref. \cite{psb1a} that the assumptions that go into proving the NW theorem does not apply to our model which uses a real scalar field that carries no definite
 baryon number, and our PSB model describes a new class of models for baryogenesis. 

To illustrate how our model provides an exception to the NW theorem, we consider a toy example which captures the main spirit of our model.
This toy model is simple, where it is straightforward to calculate baryon asymmetry obtained in the two--body decays of a real scalar field.
Our explicit calculation of $\epsilon_B$ shows that it arises in order $\alpha^2_B$ through loop diagrams that utilize $B=0$ vertices.
The loop couplings however violate ``flavor'', as will be demonstrated below.

We start with the following toy interaction Lagrangian involving a real scalar field $X$ which does not carry baryon number.
A complex scalar field $Y$, which also has $B=0$ is introduced, to mimic the effects of the $W^\pm$ gauge boson loop of our model. These fields  interact with complex bosonic fields $f_i$ with baryon number as follows (our argument also applies to the case when $f$ fields are fermionic,
but for definiteness in our calculation we take them to be bosonic): fields with same baryon number (say $B=1$) are $f_1,f_3$ and those with $B=0$ are $f_2,f_4$. The fields $(f_1,f_3)$ and $(f_2,f_4)$ can be assumed to belong to two different flavor states. When $X$ particles decay, they will generate baryons as well as anti-baryons since they produce both  $f^*_1 f_2$ and ${f}_1 f^*_2$ in their decay (and similarly for the $f_{3,4}$). The question then is:  will the $X$ decays to the two final states exactly cancel? We find below that they do not. To proceed with our proof, we start with the interaction Lagrangian
\begin{eqnarray}
{\cal L}_I=g_1 X f^\dagger_2 f_1 + g_2 X f^\dagger_4 f_3 + g_3 Y f^\dagger_3 f_1 + g_4 Y f^\dagger_4 f_2 +{\rm h.c.} \, ,
\label{int}
\end{eqnarray}
where the couplings $g_i$ have dimension of mass. 
It can be verified that not all couplings $g_i$ can be made real by field redefinitions, and one phase will survive.  Thus $C\!P$ is explicitly
violated in the Lagrangian (\ref{int}).  $X$ being a real field with no definite baryon number implies that Eq. (\ref{int}) also violates $B$.
The masses of these scalars have the form
\begin{equation}
{\cal L}_{\rm mass} = \frac{1}{2}M_X^2 X^2 + M_Y^2 Y^\dagger Y + \sum_{i=1}^4 m_i^2 f_i^\dagger f_i~.
\end{equation}
Note that there is no flavor mixing in the masses of $f_i$.  This is in fact a natural consequence of a $Z_2 \times Z_2'$ symmetry present
in the model.  The charges under this symmetry are shown in Table \ref{table_charge}.

\begin{table}[h!]
\begin{tabular}{c|c|c|c|c|c|c} \hline\hline
~ & ~$f_1$~ & ~$f_2$~ & ~$f_3$~ & ~$f_4$~ & ~$X$~ & ~$Y$~ \\ \hline
$Z_2$ & $-$ & $+$ & $-$ & $+$ & $-$ & $+$ \\ \hline
$Z_2'$ & $-$ & $-$ & $+$ & $+$ & $+$ & $-$ \\ \hline\hline
\end{tabular}
\caption{$Z_2 \times Z_2'$ charges of various fields in our toy model.} \label{table_charge}
\end{table}

We assume that $M_Y \gg M_X$ so that in the early universe, by the time $X$ particles decay,  $Y$ particles have decayed away. There are two baryon number violating final states in $X$-decay: $X \rightarrow f_1^* + f_2$ and $X \rightarrow f_3^*+ f_4$ and we must add up both the contributions.
These final states have $B = -1$, while the decays $X \rightarrow f_1 + f_2^*$ and $X \rightarrow f_3+ f_4^*$ have $B=+1$.  The net
baryon asymmetry in $X$ decays is defined as
\begin{equation}
\epsilon_B = \frac{\Gamma(X \rightarrow f_1+f_2^*) + \Gamma (X \rightarrow f_3 + f_4^*) - \Gamma(X \rightarrow f_1^*+f_2) - \Gamma (X \rightarrow f_3^* + f_4)}{\Gamma(X \rightarrow f_1+f_2^*) + \Gamma (X \rightarrow f_3 + f_4^*) + \Gamma(X \rightarrow f_1^*+f_2) + \Gamma (X \rightarrow f_3^* + f_4)}~.
\end{equation}

\begin{figure}[h!]
\centering
\includegraphics[width=5cm]{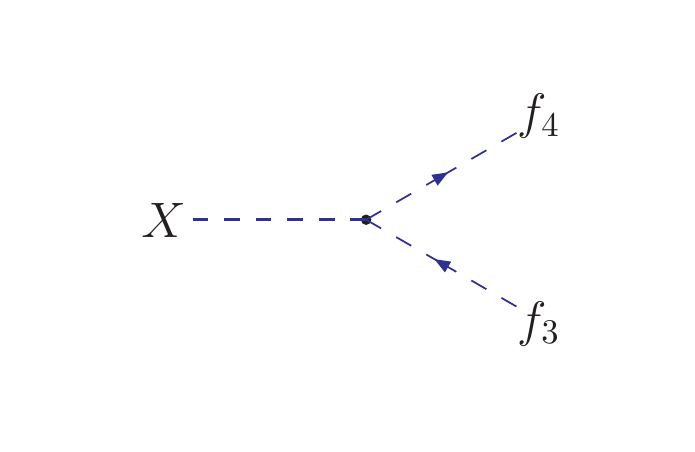}
\hspace{0.2cm}
\includegraphics[width=5cm]{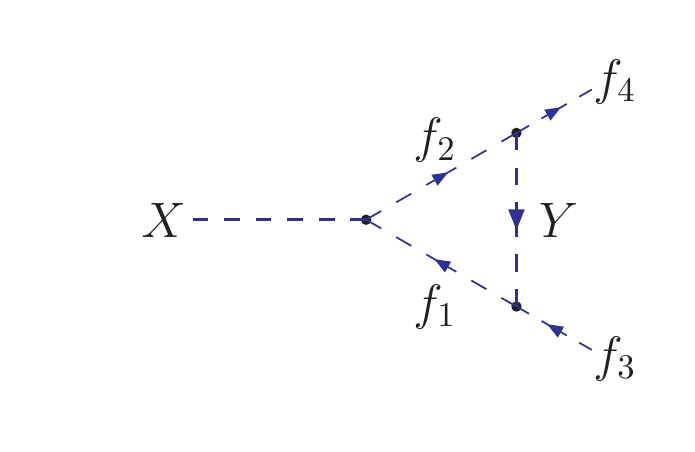}
\caption{The tree and one-loop diagram for the $X$ decay in our toy model.}
\label{fig10}
 \end{figure}
The interference of tree-level decays of $X$ with one-loop vertex corrections do lead to a net baryon asymmetry $\epsilon_B$.  
In Fig.~\ref{fig10} we show the tree level diagram and the one-loop correction which utilizes the $B$-conserving vertex of $Y$.  
The wave function correction diagrams do not generate any $C\!P$ asymmetry in this model. 
A straightforward calculation shows
(in the limit of $M_Y \gg M_X, m_i$) that
\begin{equation}
\epsilon_B = \frac{{\rm Im}(g_1^* g_2g_3 g_4^*)}{4 \pi  (|g_1|^2 + |g_2|^2)\,M_Y^2}\left[I(m_1^2,m_2^2) - I(m_3^2,m_4^2)\right]
\label{epsilon}
\end{equation}
where
\begin{equation}
I(m_a^2,m_b^2) = \sqrt{1-\frac{2(m_a^2+m_b^2)}{M_X^2} + \frac{(m_a^2-m_b^2)^2}{M_X^4}}\,\Theta\left(1-\frac{(m_a+m_b)^2}{M_X^2}\right)~.
\label{func}
\end{equation}
The $\Theta$ function signifies the absorptive part of the loop diagram.  It is clear from Eqs. (\ref{epsilon}) and (\ref{func}) that the
baryon asymmetry is non-vanishing, even though the loop diagram utilized the $B$-conserving vertex of $Y$ boson.  
The $B$ violating couplings of the model are $g_1$ and $g_2$, while $g_3$ and $g_4$ are $B$-conserving.  Our result is then
that $\epsilon_B \propto \alpha_B^2$ (in the notation of NW) and non-vanishing. The contributions from
$f_1$ and $f_2$ tend to cancel those from $f_3$ and $f_4$, but since these particles have distinct masses, there is a residual
$\epsilon_B$.  This induced $\epsilon_B$ is as a result of flavor, since it is the mass difference of flavor states that causes it. We emphasize that this is a complete calculation of $C\!P$ asymmetry in the toy model, since the only diagram
that contributes to $\epsilon_B$ is the vertex correction diagram in Fig. ~\ref{fig10}. Also this is not a gauge model so that there are no issues of gauge invariance. 
A general proof that there are exceptions to the NW theorem is presented in Ref.~\cite{gandhi} using $C\!PT$ and unitarity arguments. Here we present an explicit model that illustrates this exceptional case.

\bibliographystyle{aipproc}   

\end{document}